\documentclass[aps,preprint,showpacs,amsmath,amssymb]{revtex4}
\usepackage{epsfig}
\usepackage{wasysym}
\usepackage{amssymb}

\begin{document}

\title{Kaon photoproduction and electroproduction near threshold%
\footnote{Talk given at {\sf Baryons'10}: International Conference on 
the Structure of Baryons, December 7-11, 2010, Osaka, Japan.}}
\author{T. Mart}
\affiliation{Departemen Fisika, FMIPA, Universitas Indonesia, Depok 16424, 
  Indonesia}
\date{\today}
\begin{abstract}
We analyze the electromagnetic production of 
$K^+\Lambda$ and $K^0\Lambda$ near their production 
thresholds by using isobar models. In the $K^+\Lambda$ 
channel we show that the model can nicely describe 
the available experimental data. In the $K^0\Lambda$ 
channel we demonstrate that the $K^0$ charge form factor
has sizable effects on the longitudinal cross section.
By extending the model up to $W=1730$ MeV, we are able 
to observe the existence of the narrow $P_{11}$ 
($J^p=1/2^+$) resonance in the kaon photoproduction 
process. It is found that the most convincing mass
(width) of this resonance is 1650 MeV (5 MeV).
\end{abstract}

\maketitle
%\newpage

\section{Introduction}
Threshold properties of meson photoproduction can 
provide invaluable information, since the production dynamics 
and the complexity level of the process can be 
substantially limited at threshold. 
Especially in the case of kaon photoproduction, for which
the threshold energy is sufficiently large to allow for
a number of nucleon resonances in the process. 
For energies currently available at nuclear physics
accelerators, such as JLab in Newport News, Spring8 in Osaka,
and MAMI in Mainz, around 15 nucleon resonance listed
in the Particle Data Book (PDG) \cite{pdg2010} should be
taken  into account \cite{Mart:2006dk}. However, if we 
limit the energy of interest only up to 50 MeV above the reaction
threshold ($W\approx 1660$ MeV), only the $S_{11}(1650)$ resonance
state may contribute to the process. 

The $S_{11}(1650)$ is 
a four-star resonance. Since all of its properties are
known, we can use the information from PDG to limit 
the number of uncertain free parameters. Therefore, the 
model can be constructed to explain experimental data more accurate
than a global model that fits experimental data in a wide range of
kinematics but tends to overlook small structures near threshold. 
An accurate model is inevitably required if we want 
to investigate the origin of these structures.
Furthermore, since kaon electroproduction provides a unique
way to study charged and neutral kaon form factors in the space-like 
region, an accurate
phenomenological model is certainly well suited for this purpose.

Another interesting and relevant topic 
is the search for the nonstrange member
of antidecuplet predicted by the chiral quark soliton model 
($\chi$QSM) \cite{diakonov}, which has a mass between 1650
and 1690 MeV, but very narrow width. In the case of 
$\pi N$ or $\eta N$ photoproduction, the corresponding 
c.m. energy is far from threshold. However, this is 
not the case for kaon photoproduction, e.g. $K^+\Lambda$, for which the 
threshold energy is 1610 MeV. Clearly, kaon photoproduction
near threshold provides the best tool for this purpose.

\section{$K^+\Lambda$ Production Near Threshold}
Recently, we have investigated photoproduction of the $K^+\Lambda$ 
at energies near its production threshold by utilizing 
an isobar model \cite{Mart:2010ch}. The background 
amplitude of the model was obtained from a series 
of tree-level Feynman diagrams, whereas the resonance 
term for the $S_{11}(1650)$ state is calculated 
from the electric multipole $E_{0+}$. Although the
number of free parameters is extremely reduced
by exploiting the available information from PDG
\cite{pdg2010}, the model can nicely describe
experimental data both in the real and virtual photon
sectors up to total c.m. energy $W=50$ MeV above the threshold.

\begin{figure}[t]
  \begin{center}
    \leavevmode
    \epsfig{figure=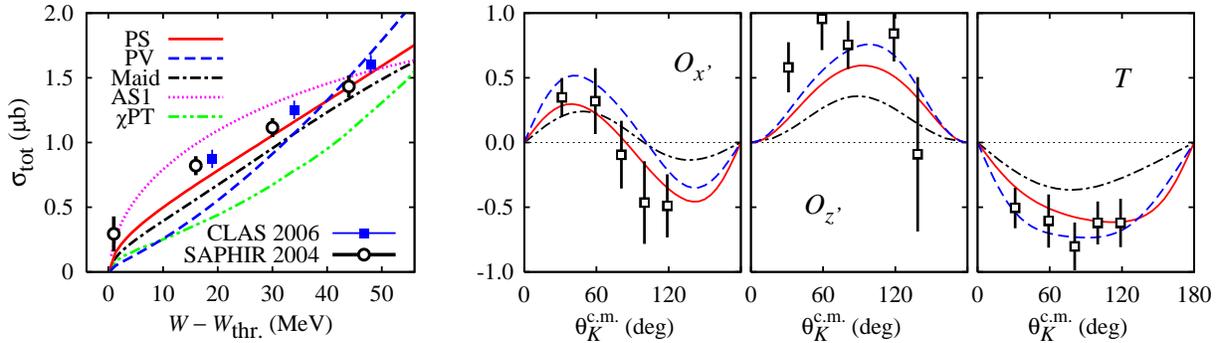,width=160mm}
    \caption{(Left panel) Comparison between total cross sections 
      calculated from the PS, PV, AS1 \cite{Adelseck:1990ch}, 
      and Kaon-Maid \cite{kaon-maid} models with the result 
      of the chiral perturbation theory \cite{Steininger:1996xw} and 
      experimental data from the SAPHIR \cite{Glander:2003jw}
      and  CLAS \cite{Bradford:2005pt} collaborations. (Right panels)
      The beam-recoil double polarization observables $O_x$, 
      $O_z$, and target asymmetry $T$ predicted
      by the PS, PV, and Kaon-Maid models \cite{kaon-maid} 
      compared with experimental data from the GRAAL 
      \cite{lleres09} collaboration. Notation of the curves is
      as in the left panel.}
   \label{fig:kpltot} 
  \end{center}
\end{figure}

It has been known that the pseudoscalar (PS) coupling can describe
kaon photoproduction data better than the pseudovector (PV) coupling.
This has been also found in our recent study, as shown in 
Fig.~\ref{fig:kpltot}. The PV result underestimates the data up to
$W=40$ MeV above threshold. The same deficiency is also shown by the
Kaon-Maid model. As shown in the figure,
prediction of chiral perturbation theory also underpredicts 
experimental data by 
about 30\% - 50\%. Interestingly, we observe that the prediction
of the PV model is very close to the prediction of the chiral perturbation 
theory up to $W=10$ MeV above threshold. This fact originates from
the small background terms of the PV model in this energy region. 
The AS1 model seems to overestimate most of
the experimental data shown. Note that we did not use these total
cross section data in our fits.

\begin{figure}[t]
  \begin{center}
    \leavevmode
    \epsfig{figure=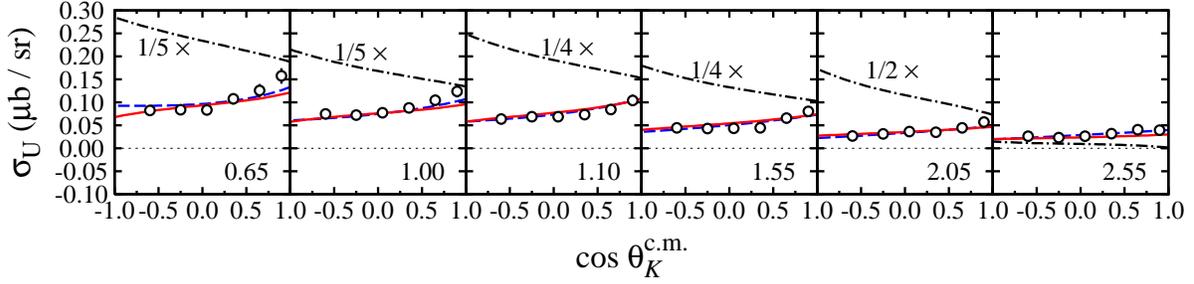,width=160mm}
    \caption{Unpolarized differential cross sections for 
      kaon electroproduction $e+p\to e'+K^++\Lambda$ as a function
      of the kaon scattering angles at $W=1.65$ GeV and
      for different values of $Q^2$. 
      Experimental data are from the CLAS collaboration
      \cite{Ambrozewicz:2006zj}. 
      Solid (dashed) lines are obtained from fit to electroproduction 
      (both photo- and electroproduction) data, whereas dash-dotted lines
      display the predictions of Kaon-Maid.}
   \label{fig:ambroz} 
  \end{center}
\end{figure}

Comparison between experimental data and 
the predicted $O_{x}$, $O_{z}$ observables as well as  
the target asymmetry $T$ is shown in the three right panels of
Fig. \ref{fig:kpltot} 
Note that the calculated polarization observables 
shown here are pure prediction. Nevertheless, 
the predicted $O_{x}$ and $O_{z}$, as well as
the target asymmetry $T$, are in perfect agreement 
with experimental data. This is true for 
both PS and PV models, although the presently available data
cannot resolve the difference between the two models.

In the case of kaon electroproduction, 
a comparison between unpolarized differential cross sections 
$\sigma_{\rm U}$ calculated from the present work and Kaon-Maid 
models \cite{kaon-maid} with experimental 
data is shown in Fig.~\ref{fig:ambroz}. In this case
it is clear that both PS and PV models can nicely reproduce 
the data and  their difference is hardly seen. 
The same result is also found in
case of the polarized structure function $\sigma_{\rm LT'}$
\cite{Mart:2010ch}. However, we notice
that Kaon-Maid is unable to reproduce the available data
in all cases and, in fact, it shows a backward 
peaking behavior for $\sigma_{\rm U}$,
in contrast to the result of experimental measurement. We
suspect that this behavior originates from the large
contributions of the $K^*$ and $K_1$ intermediate states.

\section{$K^0\Lambda$ Production Near Threshold}
By using isospin symmetry and appropriate coupling constants
obtained from PDG \cite{pdg2010}, Kaon-Maid \cite{kaon-maid}, 
as well as Pion-Maid \cite{Drechsel:2007if}, 
 the isobar model developed for the 
$K^+\Lambda$ channel in the previous 
section can be easily extended to predict
observables in the $K^0\Lambda$ channel
\cite{mart_k0lambda}. The $K^0\Lambda$
channel is of interest since in the photoproduction case
the $K^0$ $t$-channel exchange cannot contribute due to
the absence of charge and spin degrees of freedom. Thus,
the reaction mechanism is somewhat simpler and one can
also investigate the effect the $t$-channel by comparing
with the $K^+\Lambda$ case.

As briefly discussed above, the $K^0\Lambda$ electroproduction 
can be utilized to investigate the effect of $K^0$ charge form factor. 
Compared with other neutral SU(3) pseudoscalar mesons, 
the neutral kaon has a unique property, i.e. it has
an electric or charge form factor. The difference 
between the strange and non-strange quark masses 
creates a non-uniform charge distribution in the $K^0$. 
Consequently, although its total charge is zero, 
the $K^0$ has an electric or charge form factor. 

\begin{figure}[t]
  \begin{center}
    \leavevmode
    \epsfig{figure=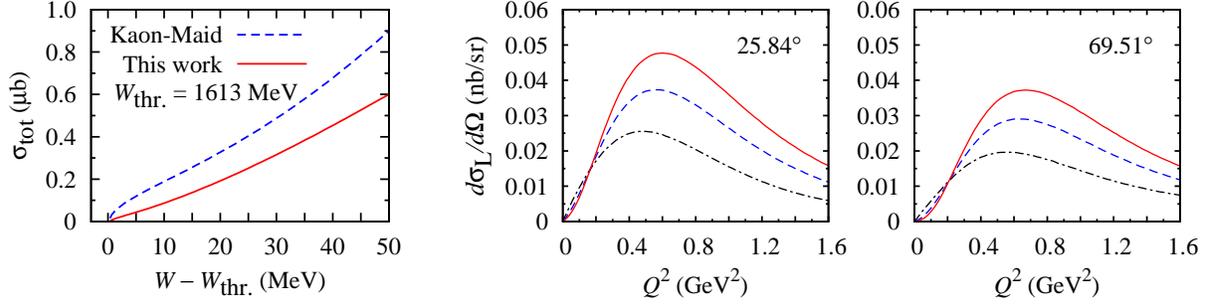,width=160mm}
    \caption{(Left panel) Comparison between total cross sections 
      of the $\gamma+n\to K^0+\Lambda$ channel 
      predicted by the Kaon-Maid \cite{kaon-maid} model and the
      present work. (Middle and right panels) Longitudinal differential cross section of the 
      neutral kaon electroproduction $e+n\to e'+K^0+\Lambda$ 
      as a function of the virtual photon momentum squared 
      $Q^2$ at $W=1.65$ GeV and for different values of the 
      kaon scattering angle. Solid lines show the calculation 
      with a $K^0$ form factor obtained in the LCQ model while 
      dashed lines are obtained by using the QMV model. The 
      dash-dotted lines are obtained from a computation with 
      the $K^0$ pole excluded.}
   \label{fig:k0ltot} 
  \end{center}
\end{figure}

The calculated $K^0\Lambda$ total cross section is shown in 
the left panel of 
Fig.~\ref{fig:k0ltot}, where prediction of the Kaon-Maid is
also displayed for comparison. It is apparent 
that even close to the threshold Kaon-Maid overpredicts 
our present calculation by a factor of about 30\%. 
The present calculation has suppressed major uncertainties,
such as unknown coupling constants, which could plague the result
of Kaon-Maid model.  Nevertheless, an experimental check of the 
$K^0\Lambda$ total cross section is still mandatory to  
help to clarify this situation.

The effects of $K^0$ form factors on the longitudinal cross section
$\sigma_{\rm L}$ are shown in the middle and right panels of 
Fig.~\ref{fig:k0ltot}, where we have
employed two relativistic quark models, the light-cone 
quark (LCQ) model~\cite{ito1} and the quark-meson vertex (QMV) 
model~\cite{buck}. It is seen from the figure 
that the effect is sufficiently large for an experimental check and 
at the forward direction the LCQ form factor raises the cross section up to 50\%. 
Since the resonance contribution is insignificant in this case, 
this phenomenon clearly originates from the dominant 
behavior of the background terms. From this figure it is also clear 
that as the scattering angle 
increases the effect slightly decreases. 
Our finding therefore corroborates the finding of 
Ref.\,\cite{Mart:1997cc}, which used the same form factors
\cite{ito1,buck}, but different 
isobar model. Therefore, experimental data with about
10\% uncertainties would be able to pin down the appropriate 
$K^0$ form factor required by the isobar model to describe 
the $e+n\to e'+K^0+\Lambda$ process.

\section{Narrow resonance in kaon photoproduction}
Although the existence
of the $\Theta^+$ has become a great puzzle nowadays, interest in
the nonstrange partner of $\Theta^+$ has increased, especially after
the finding of a narrow $P_{11}$ nucleon resonance 
effect in the $\eta$ photoproduction off the neutron 
\cite{kuznetsov}. Since the finding of this resonance, a number
of theoretical and experimental studies has been devoted for
$\eta$ photoproduction off the neutron.

The $\chi$QSM model predicts that the $P_{11}$
resonance with $J^P=1/2+$ decays mostly to $\eta N$ channels,
whereas the branching ratios to the $\pi N$ and $K\Lambda$ channels
are predicted to be comparable  \cite{diakonov}. In the $\pi N$ sector 
there is only one notable
study of this resonance \cite{igor}. In this study 
the narrow $P_{11}$ mass is obtained from $\pi N$ data
by using a modified partial wave analysis (PWA), 
since the standard PWA analysis can miss narrow
resonances with $\Gamma<30$ MeV. The changes in the total
$\chi^2$ were scanned in the range of resonance mass between 1620 to
1760 MeV after the inclusion of this resonance in the $P_{11}$
partial wave. A clear effect was observed at 1680 MeV and
a weaker one was detected at 1730 MeV. The same result 
was always obtained although the total width was varied
between 0.1 and 10 MeV and branching ratio was also
varied between 0.1 and 0.4.

To our knowledge there has been no attempt to study this
resonance by utilizing kaon photoproduction, although  
kaon photoproduction could offer a new arena for 
investigating this problem due to the explicit presence
of strangeness in the final state. In view of this we
decide to scan the changes in
the total $\chi^2$ after including a $P_{11}$ narrow 
resonance with the variation of the resonance mass, width,
and $K\Lambda$ branching ratio. Such a procedure is apparently 
suitable for kaon photoproduction, since the cross 
sections are relatively much smaller than in the
case of $\pi N$ or $\eta n$, whereas the experimental 
error bars are in general relatively larger. As 
we can see in the next section, it is difficult
to observe a clear structure in the cross sections
at the energy of interest.

\begin{figure}[t]
  \begin{center}
    \leavevmode
    \epsfig{figure=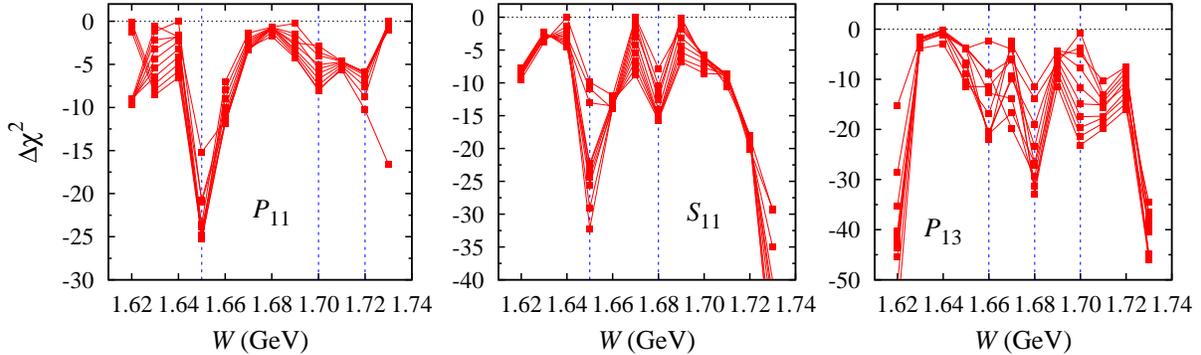,width=160mm}
    \caption{Change of the $\chi^2$ in the fit of kaon photoproduction
	data due to the inclusion of the
	$P_{11}$ (left panel), $S_{11}$ (middle panel), and
	$P_{13}$ (right panel) resonances with masses scanned from 1620 
	to 1730 MeV (step 10 MeV) and $\Gamma_{\rm tot.}$ taken from
	1 to 10 MeV (step 1 MeV) for the $K\Lambda$ branching
	ratio 0.2. The vertical lines indicate the possible
	resonance  masses.}
   \label{fig:scan_narrow} 
  \end{center}
\end{figure}

Since we shall scan the resonance mass up to 1730 MeV, we need
to extend our model explained in the previous sections up to
this energy regime. This has been performed with the addition
of the $D_{15}(1675)$, $F_{15}(1680)$, $D_{13}(1700)$, $P_{11}(1710)$, 
and $P_{13}(1720)$ resonance states \cite{mart_narrow}.

The left panel of 
Fig. \ref{fig:scan_narrow} displays the result of $\chi^2$ changes
($\Delta\chi^2$) after the inclusion of an extra $P_{11}$ resonance 
with the mass scanned
from 1620 to 1730 MeV, where the total width  $\Gamma_{\rm tot.}$
is varied from 1 to 10 MeV
with 1 MeV step. We have also investigated the variation of the 
$K\Lambda$ branching ratio $\Gamma_{K\Lambda}$ and 
found that the best $\chi^2$ would be obtained 
if we used $\Gamma_{K\Lambda}=0.2$ and $\Gamma_{\rm tot.}=5$ MeV. 
In the left panel 
of Fig. \ref{fig:scan_narrow} we can see that three minima 
appear at $m_{N^*}=1650$, 1700, and 1720 MeV. 
Nevertheless, the minimum $\Delta\chi^2$ at
$m_{N^*}=1650$ MeV seems to be the most convincing one. 

The middle panel of Fig. \ref{fig:scan_narrow} exhibits 
the changes of the $\chi^2$ if we replace the 
$P_{11}$ narrow resonance with an $S_{11}$ ($J^p=1/2^-$) resonance. 
Obviously the same minimum at $m_{N^*}=1650$ MeV
is retained, but a new one clearly appears at 
$m_{N^*}=1680$ MeV. The appearance of the minimum $\Delta\chi^2$
at $m_{N^*}=1650$ MeV in the left and middle panels 
of Fig. \ref{fig:scan_narrow}
indicates that a real structure really exists at this energy point,
although it is hardly seen in experimental data. 
However, the fact that both $S_{11}$ and $P_{11}$ could generate
this minimum means that a $J^p=1/2^-$
narrow resonance is also possible in the kaon photoproduction process.

Finally, the right panel of Fig. \ref{fig:scan_narrow}
displays the changes of the $\chi^2$ 
if we include a $P_{13}$ ($J^p=3/2^+$) resonance 
instead of an $S_{11}$ or a $P_{11}$ state. Surprisingly, the
minimum at 1650 MeV almost vanishes and a clear minimum at
1680 MeV, as in the case of the $S_{11}$, appears. Besides that, 
we also observe two weaker minima at 1660 and 1700 MeV. However,
the minimum at 1680 MeV is interesting in this case, since 
the possibility that the structure found in the $\eta$ photoproduction
off a neutron can be explained by a $P_{13}$ resonance has been
discussed in Ref.~\cite{Kuznetsov:2008hj}. In fact, the most convincing
result with the smallest $\chi^2$ would be obtained if one 
used a $P_{13}(1685)$ state instead of a $P_{11}$ \cite{Kuznetsov:2008hj}. 
Unfortunately, there is a number of meson 
photoproduction thresholds around 1685 MeV. 
Therefore, unless we could suppress 
the threshold effects at this energy point, further discussion of 
the $P_{13}(1685)$ would be meaningless at this stage.

\begin{figure}[t]
  \begin{center}
    \leavevmode
    \epsfig{figure=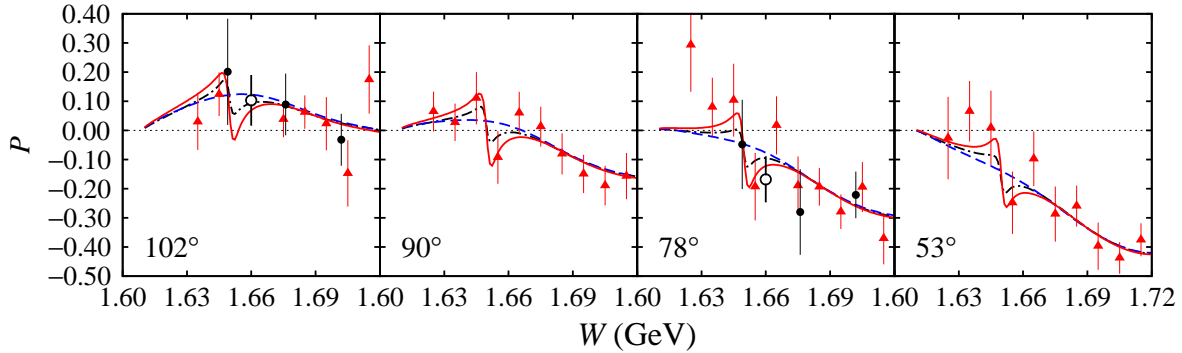,width=160mm}
    \caption{$\Lambda$ recoil polarization without the inclusion
	of the narrow resonance (dashed lines), with the inclusion
	of an $S_{11}$ resonance (dash-dotted lines) and 
	a $P_{11}$ resonance (solid lines) for different values
	of kaon scattering angles. Experimental data are from
        SAPHIR \cite{Glander:2003jw} (open circles), 
        GRAAL \cite{lleres09} (solid circles), and 
	CLAS \cite{McCracken:2009ra} (solid triangles)
	collaborations.}
   \label{fig:polar_narrow} 
  \end{center}
\end{figure}

It is obviously important to know which data are really responsible
for the minimum at 1650 MeV. For this purpose, we have scrutinized 
contributions of individual data to the $\chi^2$ in our fits
and found that this minimum originates mostly from the $\Lambda$ recoil 
polarization data as displayed in Fig.~\ref{fig:polar_narrow}.
From this figure we can see that there exists a dip at $W\approx 1650$ MeV
in the whole angular distribution of data. It is also
apparent that both $P_{11}$ and $S_{11}$ states can nicely reproduce
the dip. Therefore, it seems to us that the recoil polarization is 
not the suitable observable to distinguish the possible states at
1650 MeV. Nevertheless, more precise recoil polarization data 
are still urgently required
in order to remove uncertainties in the position of the dip.

We have also investigated total and differential cross sections, 
as well as other polarization observables
such as target and photon asymmetries, double polarization
observables $O_{x'}$, $O_{z'}$, $C_x$, and $C_z$. It is found
that total cross section and double polarization
observables $O_{x'}$, $O_{z'}$ are the appropriate observables
to determine the quantum number and parity of this narrow
resonance \cite{mart_narrow}. 

\section{Summary and Conclusion}
We have investigated the electromagnetic production of 
$K^+\Lambda$ and $K^0\Lambda$ by means of isobar models.
It is found that by using very limited free parameters we
can explain all $K^+\Lambda$ experimental data nicely. By using
isospin symmetry and appropriate parameters obtained from
PDG and Maid we predict the $K^0\Lambda$ observables and
demonstrate the possibility of investigating the $K^0$ 
form factor in the neutral kaon electroproduction.

We have also observed the existence of the $J^p=1/2^+$
narrow resonance, the nonstrange member of antidecuplet
baryons predicted by the $\chi$QSM, in kaon
photoproduction off a proton. We 
found the most promising mass and width of this resonance are 
1650 MeV and 5 MeV, respectively.
This finding is observed to be model independent
and could be distinguished from the $J^p=1/2^-$ and
$3/2^+$ resonances, provided that more precise kaon photoproduction
data were available. Furthermore, our conclusion does not
change with the variation of the total width and $K\Lambda$
branching ratio of the resonance. Although the mass 
of the resonance obtained in our calculation (1650 MeV) 
is slightly different from those obtained from 
the $\pi N$ and $\eta N$ reactions, such a mass 
corroborates the  
calculation utilizing the Gell-Mann-Okubo 
rule without mixing between the lower-lying nucleonlike 
octet with the antidecuplet. 

\section*{Acknowledgment}
The author thanks Igor I. Strakovsky for 
fruitful discussions on the search of narrow nucleon resonances.
Supports from the University of Indonesia
and the Competence Grant of the Indonesian 
Ministry of National Education are gratefully
acknowledged.

\end{document}